\documentclass[10pt]{iopart}
\usepackage{multirow}
\usepackage{graphicx}
\usepackage{hyperref}
\hypersetup{colorlinks,allcolors=black}
\usepackage{float}

\usepackage[
backend=bibtex,
sorting=none,
citestyle=nature]{biblatex}
\bibliography{ref}

\usepackage{booktabs}

\usepackage{array}
\newcolumntype{P}[1]{>{\raggedright\arraybackslash}p{#1}}

\begin{document}

\title[Dynamically Reprogrammable Stiffness in Gecko-Inspired Laminated Structures]{Dynamically Reprogrammable Stiffness in Gecko-Inspired Laminated Structures}

\author{Kai Jun Chen$^{1+}$, Maria Sakovsky$^{1*}$}
\address{$^1$ Department of Aeronautics and Astronautics, Stanford University, Stanford 94305, USA}
\ead{$^+$ckaijun@stanford.edu, $^*$sakovsky@stanford.edu}

\vspace{10pt}
\begin{indented}
\item[] Nov 2023
\end{indented}

\begin{abstract}
Adaptive structures are of interest for their ability to dynamically modify mechanical properties post fabrication, enabling structural performance that is responsive to environmental uncertainty and changing loading conditions. Dynamic control of stiffness is of particular importance as a fundamental structural property, impacting both static and dynamic structural performance. However, existing technologies necessitate continuous power to maintain multiple stiffness states or couple stiffness modulation to a large geometric reconfiguration. In this work, reversible lamination of stiff materials using Gecko-inspired dry adhesives is leveraged for bending stiffness control. All stiffness states are passively maintained, with electrostatic or magnetic actuation applied for $\sim$1s to reprogram stiffness. We demonstrate hinges with up to four passively maintained reprogrammable states decoupled from any shape reconfiguration. Design guidelines are developed for maximizing stiffness modulation. Experimentally, the proposed method achieved a stiffness modulation ratio of up to 14.4, with simulations showing stiffness modulation ratios of at least 73.0. It is anticipated that the stiffness reprogramming method developed in this work will reduce energy requirements and design complexity for adaptation in aerospace and robotics applications.
\end{abstract}


\section{Introduction}

The classic paradigm of structural design yields fixed mechanical properties representing a compromise in performance across various operating conditions. More recently, adaptive structures have focused on dynamically reprogramming mechanical performance of engineering structures \cite{Xia2022,Kuder2013} -- that is, actively changing mechanical response post fabrication. This approach enables structural performance tailored to each operating condition independently. Adaptive structures promise better response to environmental uncertainty and changing loading conditions. Dynamic reprogramming of the stiffness of structural elements is of particular interest due to its impact on a broad range of mechanical performance indices. For example, it can be used to adapt the dynamic response of structures through modification of resonant frequencies~\cite{Gavin1999,Deng2007,Tao2022}. The temporary reduction of stiffness is also advantageous for low-energy shape adaptation in load-carrying structures including morphing wings~\cite{Barbarino2011,Kuder2013}, deployable and adaptive space structures~\cite{Pellegrino2015, Testoni2021}, and robotic grippers~\cite{Firouzeh2017,Stabile2022}.

Dynamic reprogramming of structural stiffness can be attained through a variety of methods. Material-based approaches leverage temperature-dependent elastic moduli. Shape memory alloys such as NiTi exhibit a temperature- and load-dependent crystal phase transformation with a higher stiffness above their Austenite transition temperature. The different phases show stiffness modulation up to a factor of 4~\cite{Cross1969,MohdJani2014}. Shape memory polymers on the other hand show a drastic reduction of the elastic modulus above their glass transition temperature, allowing stiffness modulation as high as two orders of magnitude~\cite{Kuder2013,Behl2007,McKnight2010}. The shape memory effect in these materials is often used for recovering large deformations in adaptive structures~\cite{Firouzeh2017,Testoni2021}.

Alternatively, stiffness reprogramming can be attained via phenomena at the structural scale including contact and instabilities. Jamming techniques use meso-scale structural changes by bringing particles or lamina in contact and leveraging friction to increase stiffness. Most commonly this is done by encasing the jammed media in airtight flexible pouches and applying vacuum \cite{Narang2018,Steltz2009,Wang2021}. Conductive layers separated by thin dielectric films enable laminar jamming via elactrostatic attraction, leading to bending stiffness variation \cite{Wang2019,Bergamini2006,DiLillo2013}. Lastly, elastic buckling of thin cylindrical shells (i.e., tape springs) or other slender structural elements leads to a large reduction in bending stiffness~\cite{Rimrott1965,Pellegrino2015,Shan2015,Chen2018}. Combined with bi-stability, which allows the buckled shape to be maintained without external force, elastic instabilities can reduce energy use in structures with reprogrammable stiffness~\cite{Arrieta2014,Diaconu2007,Jeon2011,Chen2021}. While tape springs can induce stiffness changes by factors as high as 400 \cite{Pellegrino2015}, it is derived from a non-linear moment-curvature relationship and thus stiffness modulation is necessarily coupled to a large shape change.

In addition to the approaches targeted at load-carrying applications, many examples of stiffness reprogramming exist in soft robotics, biomedical, and other disciplines and have been the subject of numerous reviews~\cite{Manti2016,SaavedraFlores2013,Xia2022}.

Regardless of the technique, a continuous energy input is required to maintain at least one of the stiffness states. For example, continuous heating is necessary to maintain the high stiffness in shape memory alloys or low stiffness state in shape memory polymers, and a persistent external load is needed in jamming methods and elastically buckled structures. The exception to this are bi-stable structures where two shapes with distinct stiffnesses can be held without external force. However, bi-stability shows high sensitivity to boundary conditions \cite{Arrieta2014,Cui2015a} and couples stiffness modulation to a large shape change, making it impractical in many applications. As the field has matured, there has also been a growing interest in combining individual reprogrammable stiffness units into large mechanical metamaterials to expand the range of mechanical behavior \cite{Chen2021,Tao2022,Haghpanah2016a,Lee2022a}. Therefore, it is crucial that all operational stiffness states are passive to enable energy-efficient reprogramming and that stiffness is decoupled from shape adaption such that individual units can be readily integrated into engineered metamaterials. 

To address this gap, a novel approach is proposed in this work based on reversible lamination of thin load-carrying layers using gecko-inspired dry adhesives. We study dynamic bending stiffness reprogramming in a hinge-like element and demonstrate that two or more stiffness states can be passively maintained for any applied displacement. At the same time, we achieve stiffness modulation on par with existing methods. A combined experimental and finite element approach is presented and performance is explained within a simple analytical framework of a sandwich structure. Section~\ref{sec:concept} presents the structural and actuation concept for the reprogrammable hinge. The experimental and numerical approaches are introduced in Section~\ref{sec:methods}. Section~\ref{sec:results} demonstrates the range of achievable stiffness reprogramming and Section~\ref{sec:discussion} discusses design considerations and compares performance to the state-of-the-art. Lastly, Section \ref{sec:conclusion} concludes the paper.

\section{Reprogrammable Stiffness Concept}
\label{sec:concept}
\subsection{Working Principle}
\begin{figure}[]
	\centering
	\includegraphics[width=0.6\textwidth]{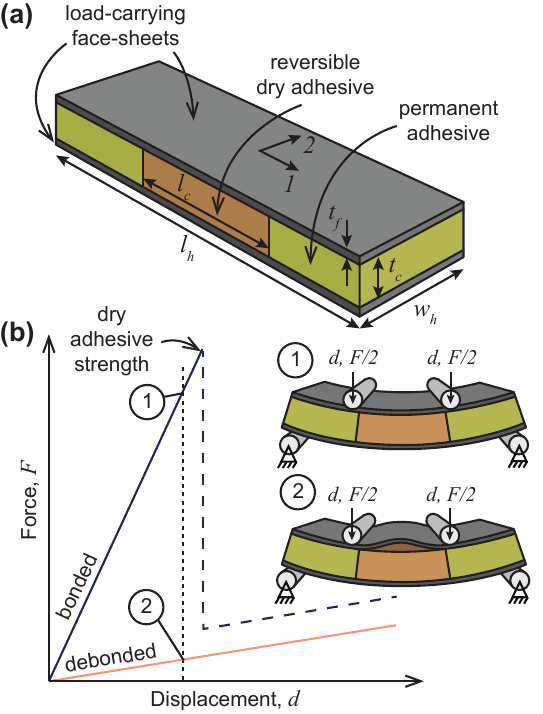}
	\caption{Reprogrammable stiffness concept using reversible lamination. (a) Schematic of hinge with reprogrammable stiffness. (b) Stiffness of hinge with load-carrying layers bonded and debonded.}
	\label{fig:concept}
\end{figure}

We take advantage of gecko-inspired dry adhesives to reversibly laminate load-carrying layers. The proposed hinge-like element with reprogrammable bending stiffness is illustrated schematically in Figure~\ref{fig:concept}(a). The hinge is composed of a functional sandwich structure at its center. Two thin load-carrying layers (`face-sheets'), with Young's modulus, $E_f$, and thickness, $t_f$, surround a soft dry adhesive core of modulus, $E_c$, and thickness, $t_c$. The dry adhesive forms van der Waals bonds with the load-carrying layers when brought into close proximity. This is achieved at the microscale through the high compliance of the dry adhesive material resulting in the formation of van der Waals bonds over a large area \cite{Wang2021a,Bartlett2012}. To maintain high absolute stiffness of the hinge, a high modulus material is selected for the load-carrying layers. A permanent stiff adhesive is used at the two ends of the hinge to limit through-thickness shear in bending.

To understand the bending stiffness reprogrammability of the proposed structure, we examine the response of the hinge in four-point bending (Figure~\ref{fig:concept}(b)). When the adhesive is bonded to the two load-carrying layers via van der Waals bonds, the structure bends as a single beam with a high bending stiffness. For sufficiently high displacements, the dry adhesive strength is exceeded and an elastic buckling instability occurs on the compression side, associated with adhesive debonding. Note that for stiffness control purposes, only the initial high stiffness region is of interest. On the other hand, when the dry adhesive is initially debonded, the two load-carrying layers bend upon their individual neutral axis and the bending stiffness is significantly reduced. The face-sheet on the compressive side will elastically buckle due to its small thickness. We refer to these two stiffness states as bonded and debonded, respectively, and note that they can both be passively maintained at any applied displacement (up to the adhesive strength). As such, external energy input is required only to bond and debond the dry adhesive and stiffness reprogramming is decoupled from shape reconfiguration. 

The stiffness ratio, $\rho$, between the bonded and debonded states can be estimated analytically from Classical Lamination Theory (CLT)~\cite{Daniel2006},
\begin{equation}
	D_{11} = \sum_{k} Q_{11}^k(z_k - z_{k-1})^3
	\label{eqn:DbCLT}
\end{equation}
where $D_{11}$ is the longitudinal bending stiffness per unit width of the hinge, following standard CLT notation, $Q_{11}^k$ is the longitudinal axial stiffness of layer $k$, and $z_k$ are the locations of the layer boundaries from the mid-plane. For the middle section of the hinge where two load-carrying face-sheets are bonded via a dry adhesive core, Equation~\ref{eqn:DbCLT} gives,
\begin{equation}
	D_{11,b} = \frac{1}{2}Q_{11}^f t_f t_c^2 \left[ 1 + 2 \frac{t_f}{t_c} + \frac{4}{3} \left(\frac{t_f}{t_c}\right)^2\right]
\end{equation}
where $c$ and $f$ superscripts refer to the core and face-sheets, respectively, and the bending stiffness contribution of the core layer is neglected since $E_c << E_f$. The debonded bending stiffness can be approximated by adding the bending stiffness of each individual layer,
\begin{equation}
	D_{11,d} = \frac{1}{6}Q_{11}^f t_f^3
\end{equation} 
where the bending stiffness contribution of the core is again negligible. The stiffness ratio is therefore only a function of the thickness of the layers,
\begin{equation}
	\rho = \frac{D_{11,b}}{D_{11,d}} = 3\left(\frac{t_c}{t_f}\right)^2 + 6\left(\frac{t_c}{t_f}\right) + 4
	\label{eqn:rho}
\end{equation}

This analysis assumes negligible transverse shear deformation in the silicone core (i.e., normals remain normal upon bending). In practice, the stiff permanent adhesive at the ends of the hinge will limit transverse shear deformation in the soft dry adhesive core but will not eliminate it entirely, particularly for high $t_c/t_f$. We also neglect the elastic buckling of the face-sheet on the compression side of the bend in the debonded state. The validity of these assumptions will be investigated in Section~\ref{sec:results}. Regardless, CLT provides a simple analytic framework to understand the bending stiffness reprogrammability.

Equation~\ref{eqn:rho} shows that for an infinitesimally thin dry adhesive core, this approach yields a stiffness ratio of 4. Increasing the dry adhesive core thickness relative to that of the face-sheets, drastically improves the stiffness variability. For example, for $t_c/t_f = 2$, we expect a maximum stiffness variation by a factor of 28. As will be investigated in Section~\ref{ssec:multiLayer}, the adhesive and load-carrying layers can be repeated to obtain more than two stiffness states and increase stiffness variation further.

Reversible lamination using dry adhesive yields an effective method of dynamic stiffness reprogrammability where all stiffness states are passively maintained, stiffness modulation and shape adaptation are decoupled, and the stiffness modulation is easily controlled through geometric parameters.

\subsection{Actuation Concepts}
In this work, we propose two methods for lightweight bonding and debonding of the dry adhesive: electrostatic and magnetic actuation. 

In the electrostatic approach, metallic face-sheets are used as both the load-carrying layers and actuation. Application of a large potential difference ($\mathcal{O}(kV)$) between the two face-sheets creates an electrostatic attractive force between them. This creates a pressure between the dry adhesive and face-sheets and is sufficient to bond the layers. Debonding cannot be achieved using this method and is instead done through external mechanical loading. 

In the magnetic approach, non-magnetic face-sheets are used. A small electromagnet realized from a coil of wire is bonded to one face-sheet and a permanent magnet is bonded to the other face-sheet. Application of a current through the coil in one direction attracts the permanent magnet and applies pressure between the dry adhesive and face-sheets, thereby bonding them. Reversal of the current repels the permanent magnet and breaks the van der Waals bonds to debond the layers. 

The two methods can theoretically achieve rapid stiffness switching, on the order of the time constant of the actuation circuit ($\mathcal{O}(ms)$)~\cite{Wang2019}. Embedding of wires in the dry adhesive to heat the layer via resistive heating was also considered. Expansion of the dry adhesive would apply the required pressure for bonding. However, this method has the disadvantage of slow actuation rates and was not pursued.

\section{Materials and Methods}
\label{sec:methods}
\subsection{Materials}

The electrostatically actuated hinge was fabricated from thin $t_f = 100~\mu m$ steel sheets (1095 spring steel) for the load carrying layers and thickness-controlled silicone sheets (MoldStar\textsuperscript{TM} 30) for the dry adhesive. An epoxy adhesive (Loctite EA E-20HP) was used to bond the face-sheets at the two ends.

In the magnetically actuated hinge, load-carrying layers were manufactured from HR40 carbon fiber and Thinpreg 513 epoxy prepreg from North Thin Ply Composites with an areal weight of 45~$g/m^2$ and fiber volume fraction $V_{f} = 55~\%$. The dry adhesive was realized from thickness-controlled silicone sheets (Bluesil RTV 3428). To enable actuation, three $10~mm \times 10~mm \times 1.2~mm$ Neodymium permanent magnets were used in conjuction with an electromagnet coil made from enamel-coated Copper wire with a diameter of 0.22~$mm$ and 80 turns. The coil was cast in epoxy to better distribute actuation pressures~\cite{schmid2022variable}.

The material properties of all constituents are summarized in Table~\ref{tab:MaterialProps}. 

\begin{table*}[h!]
	\small
	\centering
	\begin{tabular}{lllllll}
		Hinge Type & Material & $E_{1}$ [GPa]& $E_{2}$ [GPa] & $G_{12}$ [GPa] & $\nu_{12}$ [-] & $V_f$ [-] \\ \hline
        \multirow{3}{*}{Electrostatic Actuation}
        & Silicone & 0.00066 & 0.00066 & 0.00022 & 0.49 & - \\
		& Steel & 206.0 & 206.0 & 147.1 & 0.3 & - \\
		& Epoxy & 2.1 & 2.1 & 1.7 & 0.38 & - \\ \hline
       \multirow{2}{*}{Magnetic Actuation}
        & Silicone & 0.0080 & 0.0080 & 0.0078 & 0.49 & - \\
		& CFRP & 206.2 & 9.1 & 3.4 & 0.26 & 0.55 \\ \hline        
	\end{tabular}
	\caption{Material properties of hinge constituents.}
	\label{tab:MaterialProps}
\end{table*}

\subsection{Hinge Fabrication}

A thickness-control process was adopted for the fabrication of all components and hinges in this study. The respective materials were cast in their uncured state between two thick stainless steel plates with a surface roughness of 1.6~$\mu m$ and a flatness of 0.03 mm per 100 mm. Precision steel spacers were used between the plates to set the desired thickness.
 
\subsubsection{Dry Adhesive}

Dry adhesives typically use micro-scale patterning on their surface with the compliance of these features critical to achieve large contact areas for van der Waals bonds~\cite{Wang2021a}. Bartlett et al. proposed that scalable dry adhesion can instead be attained without patterning by curing silicone on a sufficiently smooth surface and using a fiber reinforcement to increase the shear strength~\cite{Bartlett2012}. We follow this strategy in this study. The adhesive strength is dominated by the normal stresses applied to the dry adhesive by the elastic buckling of the face-sheet on the compressive side of the fold. As a result, the fiber-reinforcement of the adhesive is omitted here.

The silicone was cast between the two plates and cured for 24~hours. To facilitate the removal of the dry adhesive after cure, the two plates were staggered and one was lined with a 50 $\mu m$ polyimide sheet (Dupont Kapton HN).

\subsubsection{Electrostatically Actuated Hinge}
\begin{figure}[h]
	\centering
	\includegraphics[width=0.6\textwidth]{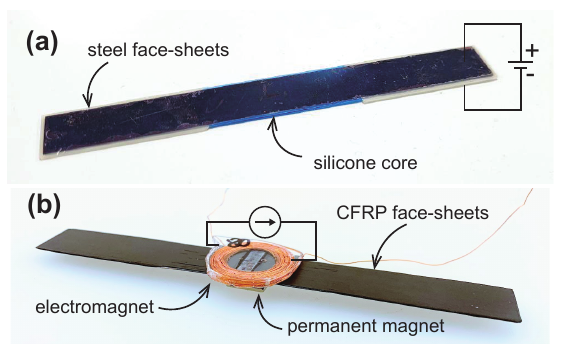}
	\caption{(a) Photograph of electrostatically actuated hinge. (b) Photograph of magnetically actuated hinge.}
	\label{fig:hinges}
\end{figure}

The steel layers were cut to size with a paper guillotine. A sharp blade was crucial to ensure that the steel remained flat after cutting. The face-sheets and silicone were stacked on the steel plate and the permanent epoxy adhesive was applied between the face-sheets with a spatula. The hinges were clamped between 10~$mm$ thick steel plates. Excess epoxy was squeezed out during cure due to the spacers and was trimmed with scissors. A photograph of an electrostatically manufactured hinge is shown in Figure~\ref{fig:hinges}(a). 

\subsubsection{Magnetically Actuated Hinge}
The magnetically actuated hinges were made from carbon fiber reinforced polymer (CFRP) prepreg. The prepreg was cut to size and stacked to the desired thickness. For simplicity, unidirectional laminates were used in this study, with the carbon fibers parallel to the hinge length. The load-carrying layers used two layers of prepreg for a total cured thickness of 96~$\mu m$ while the number of CFRP layers in the ends was varied to match the silicone adhesive thickness. The prepreg was stacked between two 10~$mm$ thick steel plates coated with release agent. The spacer thickness was selected to keep the resin bleed between 2 -- 10\% by weight of the composite. The assembly of prepreg and silicone core was cured in an autoclave using the manufacturer recommended cure cycle (2 hours at 120$^o$C).

Once cured, the actuation was boded to the hinge using adhesive tape (Tesa 05338). The electromagnet was attached at its center using a single 3.5~mm wide strip across the width of the hinge while the permanent magnets used two strips at their ends. This attachment of the actuation was found to best distribute pressure from the actuation and produced the most repeatable results~\cite{Schmid2022}. A photograph of a magnetically actuated hinge is shown in Figure~\ref{fig:hinges}(b).

\subsubsection{Geometry of Hinge Prototypes}
The thickness of all manufactured prototypes is summarized in Table~\ref{tab:hingeGeom}. All hinges had a length of $l_h = 120~mm$ with a $l_c = 42~mm$ long silicone core. The width was $w_h = 13~mm$ for both the CFRP and steel hinges. For the steel hinge, silicone adhesive thickness values were measured using a micrometer and the load-carrying layer thicknesses are nominal values according to manufacturer data \cite{McMaster9014K11}. For the CFRP hinge, all values were measured using optical microscopy (Keyence VHX-6000). All prototypes had two load-carrying layers with a single adhesive core with the exception of one steel prototype that repeated the layers for a total of three load-carrying and two dry adhesive layers. The latter demonstrated three programmable stiffness states.

\begin{table}[h]
	\small
	\centering
	\begin{tabular}{lllll}
		Hinge Type & $t_c$ [$\mu m$] & $t_f$ [$\mu m$] & $t_c/t_f$ \\ \hline
  		CFRP & $289.0 \pm 5.6$ & $90.9$ & 3.18 \\ \hline
        \multirow{4}{*}{Steel}
        & $219.3 \pm 5.1$ & $100.0$ & 2.19 \\
        & $323.6 \pm 8.6$ & $100.0$ & 3.24 \\
        & $432.2 \pm 9.1$ & $100.0$ & 4.32 \\
        & $497.9 \pm 6.3$ & $100.0$ & 4.98 \\ \hline
        Steel (3 Layer)  & $330.9 \pm 8.1$ & $100.0$ & 3.31 \\\hline
	\end{tabular}
	\caption{Geometry of hinges in study. The error indicates the standard deviation of the measurements. The 3 layer steel hinge has 2 dry adhesive cores and 3 load-carrying layers.}
	\label{tab:hingeGeom}
\end{table}

\subsection{Experimental Characterization}
The hinge stiffness was characterized using four-point bending (illustrated schematically in Figure~\ref{fig:concept}(b)). For the electrostatically actuated hinges, tests were conducted on an Instron 68TM-50 universal testing machine (UTM) with a 100N load cell. A load span of 45~$mm$ and a support span of 90~$mm$ was used. For the magnetically actuated hinges, tests were conducted on a Zwick Roell Z005 AllRoundLine UTM with a 5 kN load cell. A load span of 52~$mm$ and a support span of 104~$mm$ was used. In both cases, the hinges were loaded at 5~$mm/min$ up to a $d = 2 mm$ displacement.

The hinges were actuated directly on the testing machine to minimize the effects of manual handling on the stiffness response. 3D printed PLA bending fixtures were used to minimize interaction with the actuation.

For the electrostatically actuated hinges, a high voltage supply (Stanford Research Systems PS350) was used to apply potential differences up to 1000$V$. For the magnetically actuated hinges, bonding and debonding was done using a DC power supply (RND 320-KA3305P) controlled using the Korad KA3005P software. A prescribed current of 5$A$ was applied to the electromagnet for 1.4$s$ to achieve actuation.  In both cases, actuation times had no impact on the stiffness response, and times as low as 1$s$ were sufficient to successfully bond the layers of the hinge. 

\subsection{Finite Element Simulations}
\begin{figure}[h]
	\centering
	\includegraphics[width=0.6\textwidth]{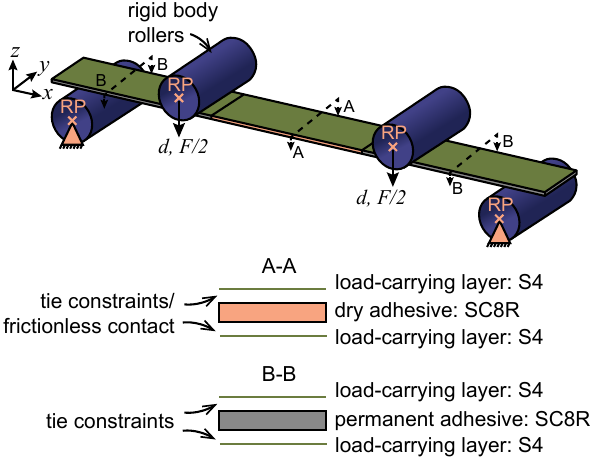}
	\caption{Schematic of finite element model.}
	\label{fig:FEM}
\end{figure}

A finite element (FE) model simulating a four point bending test of the hinge was implemented in Abaqus/Standard (Figure~\ref{fig:FEM}). The model captures the hinge bending stiffness in the bonded and debonded states and does not attempt to model the adhesive strength. The schematic in Figure~\ref{fig:FEM} shows a hinge with two load-carrying layers and a single dry adhesive layer as in Figure~\ref{fig:concept} and in the majority of physical prototypes. The FE model can also account for multi-layer hinges with an arbitrary number of dry adhesive layers as will be discussed in Section~\ref{ssec:multiLayer}.

The hinge was modeled as a laminate with two layer types (load-carrying layers and adhesive) with different interaction properties between the layers used to capture the stiffness responses in the bonded and debonded states. The load carrying elements were modeled by their mid-planes using S4 quadratic shell elements with their material properties defined using an orthotropic material to capture all material options in this study. The adhesive layer was modeled using SC8R reduced integration continuum shell elements. This is important to capture the correct response of the relatively thick and soft silicone by allowing the elements to change thickness and accounting for transverse shear deformation~\cite{Simulia2021}. The adhesive layer was partitioned to assign separate material properties to the silicone dry adhesive and permanent adhesive regions. A mesh size of 1~$mm$ was used for all elements, determined using a mesh convergence study.

Tie constraints were used to simulate permanent bonding between the load-carrying layers at their two ends. To simulate the bonded stiff state, tie constraints were also implemented between the load-carrying layers and the dry adhesive regions. To capture the debonded soft state, frictionless tangential contact and hard normal contact were used between the two instead.

All load-carrying layers that elastically buckle contained a small imperfection to mimic the imperfection-sensitive buckling behavior of thin shells~\cite{Brush1975}. This was implemented through a 30~$\mu m$ amplitude geometric imperfection in the center section of the layer. This method of modeling the imperfection sensitivity was most physical as the manufactured hinges were not perfectly flat.

Bending was applied using four rigid rollers with a 5.0~$mm$ radius, matching experiments. The motion of each roller was controlled by coupling all of its degrees of freedom to a reference point (RP). The boundary conditions used to apply bending to the hinge are illustrated in Figure~\ref{fig:FEM}. Frictional contact with a penalty formulation and a friction coefficient of 0.2 was used to model the tangential interaction between the rollers and the hinge. Hard contact was used for normal behavior. 

A geometrically non-linear dynamic implicit analysis was used to solve for the hinge deformation. The quasi-static option was enabled and the loading rate was chosen to limit the kinetic energy at all simulation increments to below 5\% of the internal energy. Through this approach, we are able to capture the quasi-static experimental response and circumvent the convergence difficulties of the static solver associated with the elastic buckling instability. An arc-length method can be used instead if the post-buckling response is of interest \cite{Riks1979}. 

\section{Results}
\label{sec:results}

\subsection{Stiffness reprogramming proof of concept}
\begin{figure*}[!htb]
	\centering
	\centerline{\includegraphics[width=1.2\textwidth]{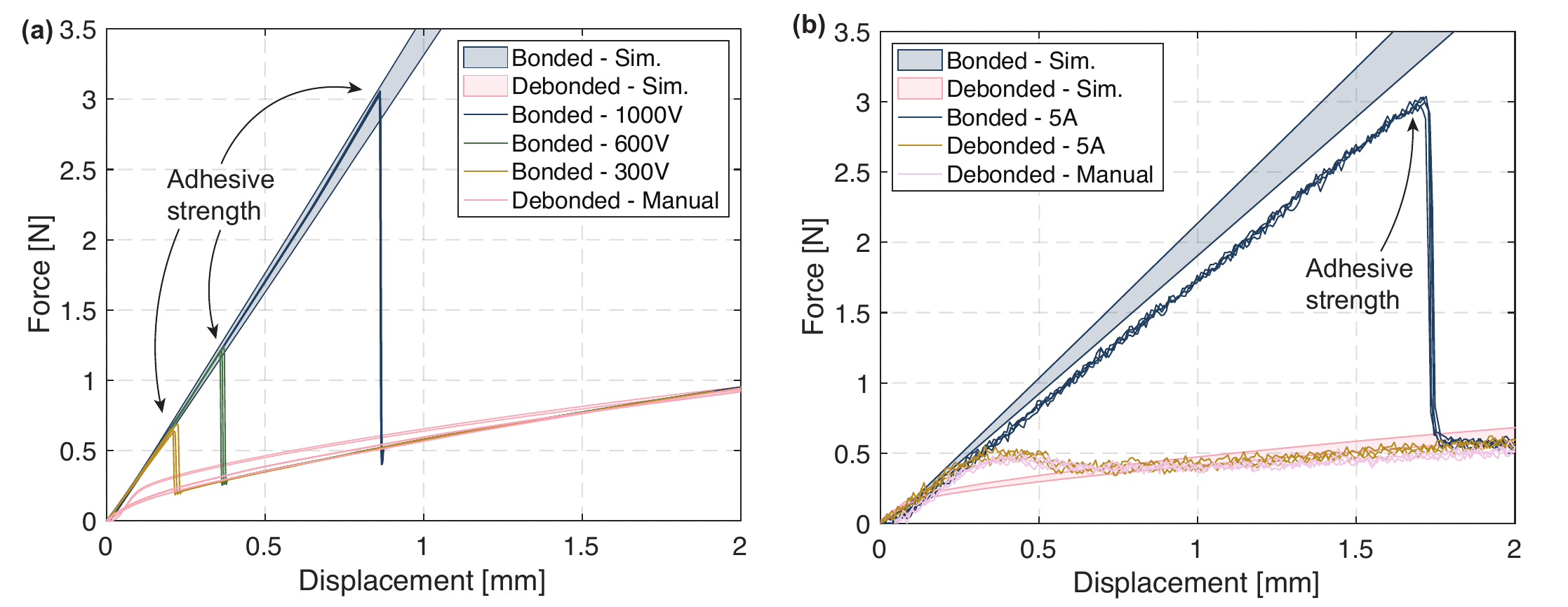}}
	\caption{(a) Stiffness reprogramming in electrostatically actuated hinge with $t_{c}/t_{f}$ = 3.24. (b) Stiffness reprogramming in magnetically actuated hinge with $t_{c}/t_{f}$ = 3.18.}
	\label{fig:proofOfConcept}
\end{figure*}

\begin{figure*}[!htb]
    \centering
    \centerline{\includegraphics[width=1.2\textwidth]{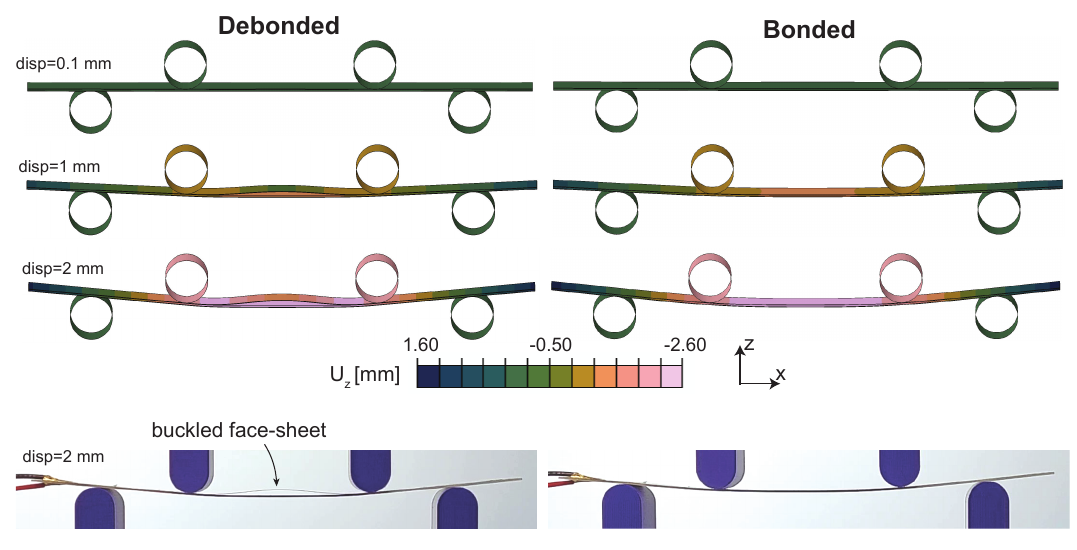}}
	\caption{Snapshots of the simulated and experimental deformation at various crosshead displacements for the bonded and debonded steel hinge with $t_{c}/t_{f}$ = 3.24. Contour is out-of-plane displacement in the z-direction.}
	\label{fig:deformation}
\end{figure*}

Stiffness modulation is demonstrated for the electrostatically actuated hinge in Figure~\ref{fig:proofOfConcept}(a). The manually debonded response of the hinge is compared to electrostatic actuation at voltages of 300~V, 600~V and 1000~V. The debonded response shows a small initial region with a higher stiffness before the buckling load of the face-sheet on the compression side is reached. Following this, the desired low stiffness linear regime can be observed. After actuation, we observe a much larger linear stiff response, until the dry adhesive strength is exceeded. The experimentally observed stiffness ratio between the linear portions of the bonded and debonded behaviors is $\rho = 8.1$. 

Three bonding trials are shown for each actuation voltage, demonstrating excellent repeatability in both bending stiffness (std. dev. = 1.3\%) and debonding load (std. dev. = 1.8\%). Notably, actuation is only applied for $\sim$1s to achieve bonding using the dry adhesive, while both stiffness states can be maintained passively. Higher actuation voltage is beneficial for increasing the adhesive strength, likely due to the formation of Van der Waals bonds over a larger area due to higher electrostatic force. However, it is observed that the adhesive strength is also sensitive to imperfections in the core thickness arising from manufacturing defects. Precise control of clamping forces when curing the epoxy ends is necessary to achieve precise contact between layers without introducing inherent compressive stresses into the core adhesive layer. 

The FE predictions for the two operating regimes are shown in the shaded regions in Figure~\ref{fig:proofOfConcept}(a). The bounds of the regions correspond to the measured standard deviation of the thickness of the dry adhesive layer. The debonded response shows good agreement with the predicted stiffness. The simulations similarly show a small higher stiffness region before elastic buckling, although this region is higher in simulations than measured. A potential explanation is  geometric imperfection in the hinge specimen, causing it to buckle more easily. The bonded prediction has excellent agreement with experimental results.

The simulated and experimental deformation of the steel hinge in its debonded and bonded states is illustrated in Figure~\ref{fig:deformation}. The buckling of the face-sheet under compression is clearly evident in the debonded case for displacements above $d = 0.1~mm$. This response is absent in the bonded case.

Stiffness reprogramming is demonstrated in the magnetically actuated hinge as shown in Figure~\ref{fig:proofOfConcept}(b). In this case, we demonstrate 3 actuation cycles and show that the bonding and debonding actuation is successful at changing the bending stiffness response of the hinge reversibly. Specifically, the debonding actuation yields the same response as a manually debonded hinge. For the magnetically actuated hinge, the measured stiffness ratio is $\rho = 12.7$. Simulated behavior agrees well with experimental measurements.  

\subsection{Control of stiffness modulation using geometry}

\begin{figure}[h]
	\centering
	\includegraphics[width=0.6\textwidth]{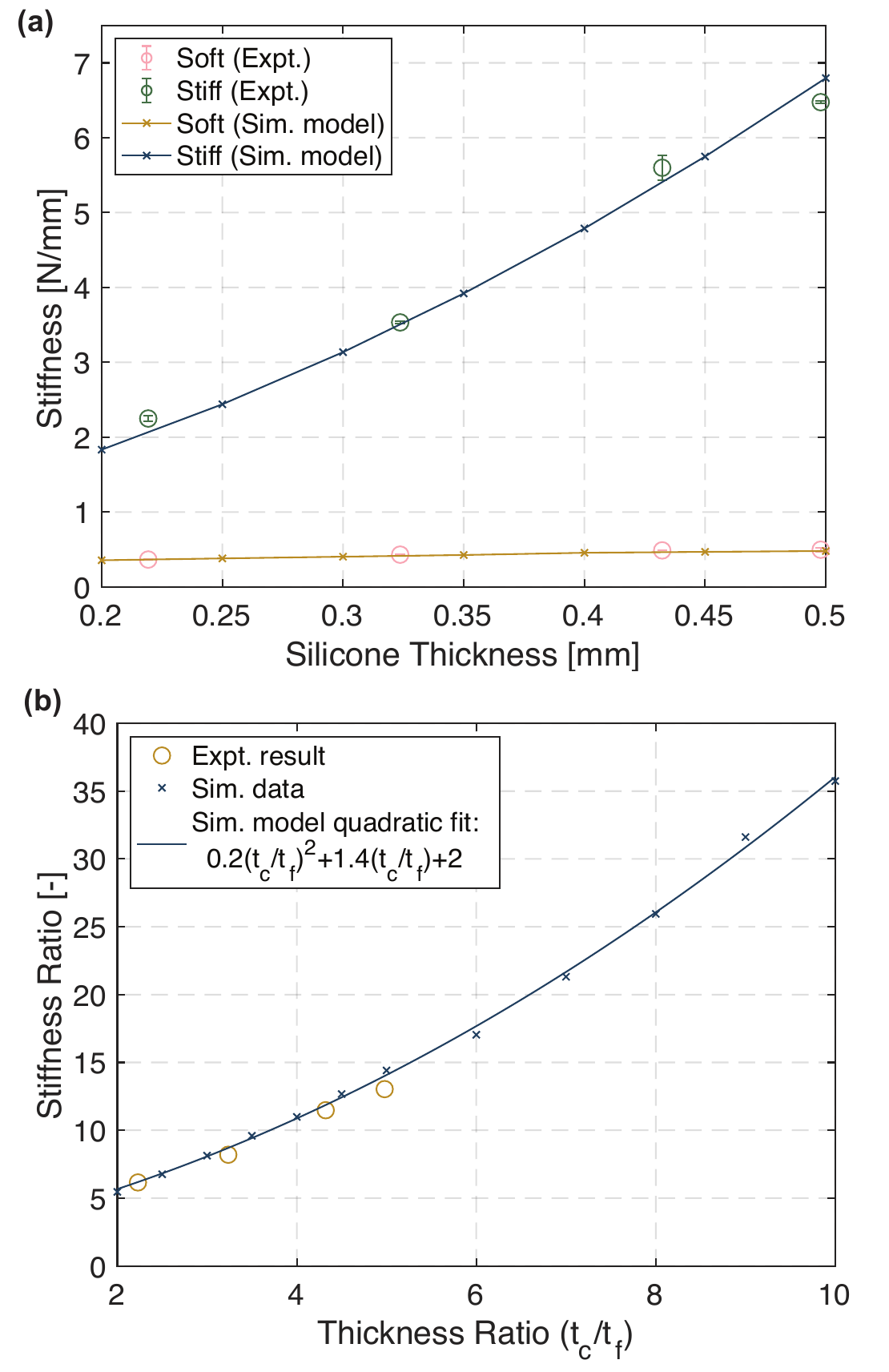}
	\caption{Control of stiffness ratio using thickness of silicone core. (a) Variation of bending stiffness for steel hinges in debonded (soft) and bonded (stiff) states.  The error bars indicate the standard deviation in experimental results. (b) Quadratic dependence on $t_c/t_f$ found in simulation, with experimental values included for comparison.}
	\label{fig:stiffnessRatio}
\end{figure}

We test several steel hinges with identical face-sheets but different silicone core thicknesses (as per Table~\ref{tab:hingeGeom}). The measured bending stiffness response of these is shown in Figure~\ref{fig:stiffnessRatio}(a) and is in good agreement with FE simulation results. Each sample is tested 3 times in each stiffness state for repeatability, with bonding and debonding done manually. Experimental stiffness values are computed from a 0.3 N pre-load up to a 1.0 $mm$ cross-head displacement. As expected, the bonded stiffness is easily increased using the core thickness, $t_c$. The experimental design with the thickest core of $t_c = 498~\mu m$ achieves a stiffness ratio of $\rho = 13.0$.

A broader range of core thicknesses in the steel hinges is simulated in Figure~\ref{fig:stiffnessRatio}(b) and is in good agreement with experimental results. A quadratic dependence of the stiffness ratio, $\rho$, on the thickness ratio $t_c/t_f$ is observed. This is as expected from Equation~\ref{eqn:rho}. However, the coefficients are much lower than what is predicted analytically leading to lower stiffness modulation. We revisit the model assumptions to understand the discrepancy in the next two sections.

\subsection{Comparison of stiffness modulation with analytic prediction}
The analytic model in Equation~\ref{eqn:rho} assumes that the rigid ends of the hinge prevent significant shearing of the soft dry adhesive core. As a result, we use CLT without an additional correction factor for transverse shear as is done for thick composites~\cite{Daniel2006}. A transverse shear correction factor of 5/6 is used in the Abaqus finite element software for thick shells and is implemented by default \cite{abaqus2014abaqus}.

\subsubsection{Deformation of dry adhesive core}
We examine the extent of transverse shear in the middle of the hinge through the longitudinal curvature, $\kappa_1$, of the shells in the bonded state. The most notable aspect is that the curvature is non-uniform in the section between the inner rollers contrary to what would be expected in a four-point bending test that applies pure bending to this section. This is indication of the effects of transverse shear. However, the deviation of the deformation from pure bending has no effect on the overall stiffness of the hinge since $E_c << E_f$. We verify through FE modelling that increasing stiffness of the dry adhesive up to 1\% of the stiffness of the load-carrying layers (i.e., dry adhesive with a modulus similar to epoxy) has no impact on the overall stiffness of the hinge. Similarly, modification of the transverse shear stiffness of the dry adhesive from 1 - 1000\% that of the computed default has no impact on hinge stiffness. This crucial observation allows us to use a relatively soft material like silicone for its functional properties without sacrificing overall stiffness. 

\subsubsection{Deformation of the buckled face-sheet}
\begin{figure}[h]
	\centering
	\includegraphics[width=0.6\textwidth]{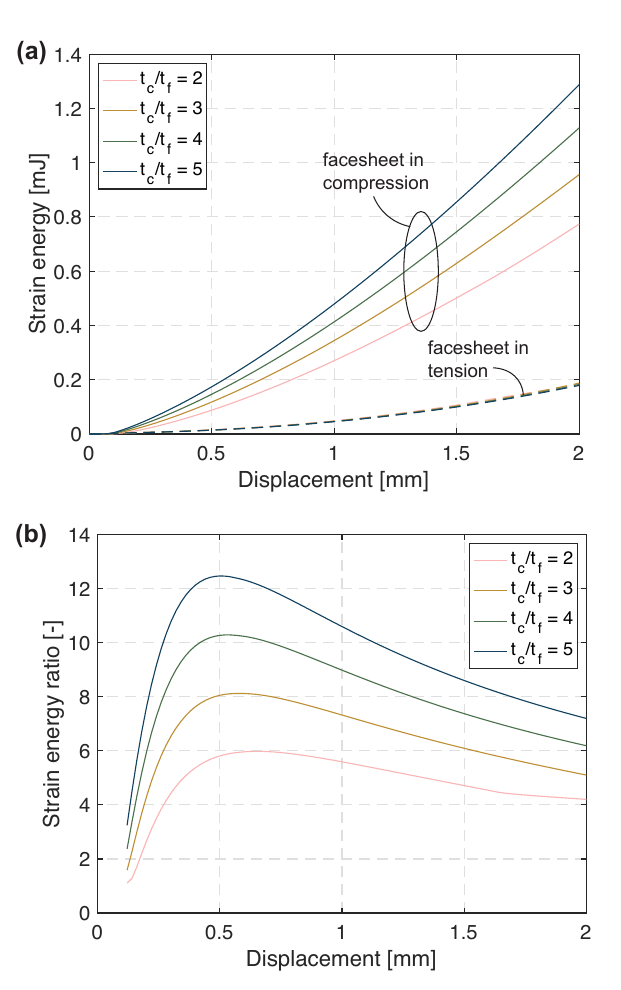}
	\caption{Strain energy of two face-sheets in the debonded state. (a) Strain energy as a function of applied displacement for several thickness ratios. (b) Ratio between the strain energy stored in the face-sheet under compression and the one in tension for several thickness ratios.}
	\label{fig:strain_energy}
\end{figure}

Above, we showed that the soft silicone core does not impact the bonded high stiffness state of the hinge. We now examine the debonded soft stiffness state in more detail to understand the obtained stiffness ratios. Figure \ref{fig:strain_energy}(a) shows the deformation strain energy in the debonded state for the load-carrying layers on the tensile and compressive sides of the bend for several thickness ratios. It is observed that in all cases the elastically buckled face-sheet on the compression side deforms to higher curvatures and therefore stores higher strain energy. The deformation of the face-sheet on the tensile side is independent of the hinge geometry (i.e., thickness ratio) as it bends upon its own neutral axis. The energy stored in the elastically buckled face-sheet increases with the thickness ratio. The further away this face-sheet is from the mid-plane of the hinge, the larger the compressive buckling force that is applied to it. In fact, this face-sheet stores as much as 12 times the strain energy of the face-sheet on the tensile side as can be seen in Figure \ref{fig:strain_energy}(b). Therefore, the elastic buckling of the face-sheet in the debonded states must be accounted for when predicting the stiffness ratio as it significantly increases the stiffness of the soft debonded state.

\subsection{Expanding the stiffness modulation behavior}
\label{ssec:multiLayer}

\begin{figure*}[h!]
	\centering
	\centerline{\includegraphics[width=1.2\textwidth]{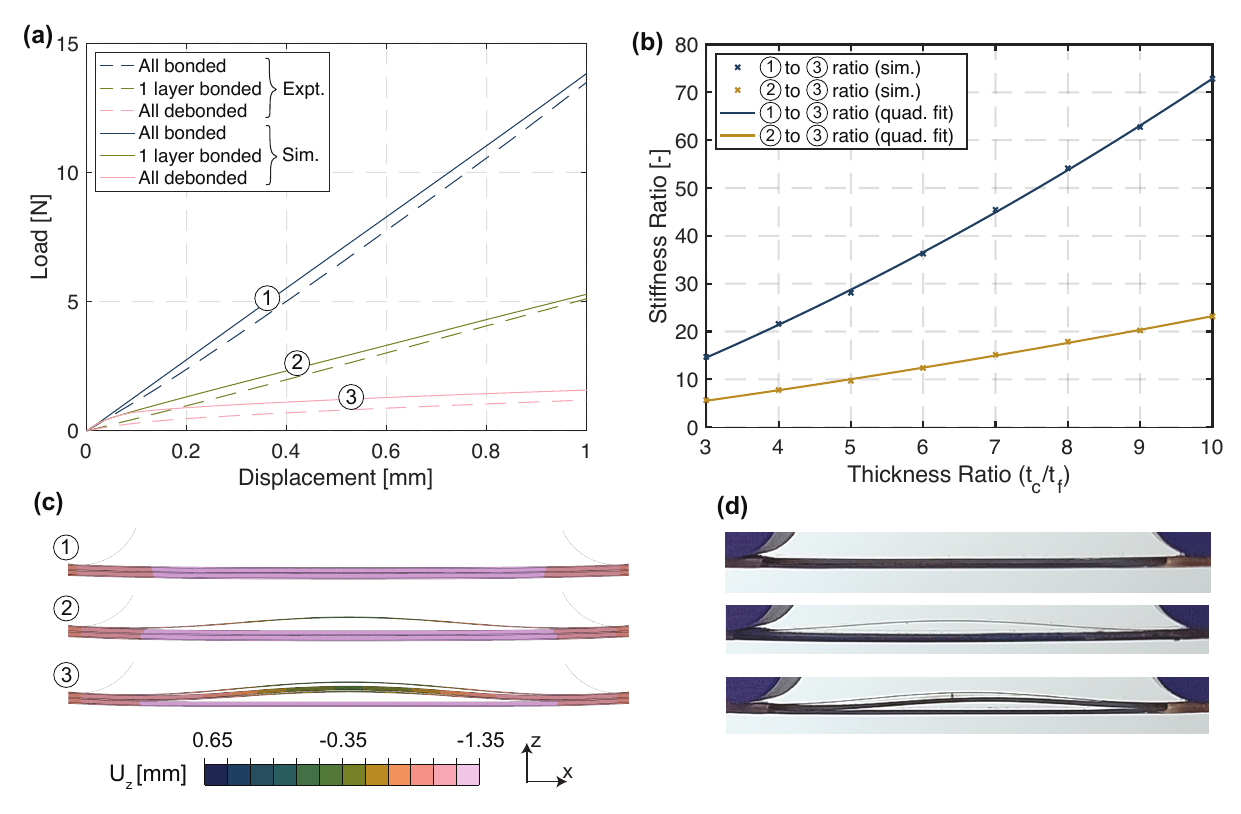}}
	\caption{Control of stiffness ratio using 3 load-carrying layers. (a) Simulation and experimental results of force-displacement response of the hinge with $t_c/t_f = 3.31$. (b) Achievable stiffness ratios of the hinge as a function of thickness ratio. (c) Snapshots of simulated and experimental deformation of force-displacement response of the hinge with $t_c/t_f = 3.31$, at $d = 1.0~mm$.}
	\label{fig:3Layer}
\end{figure*}

\begin{figure}[h!]
	\centering
	\includegraphics[width=0.6\textwidth]{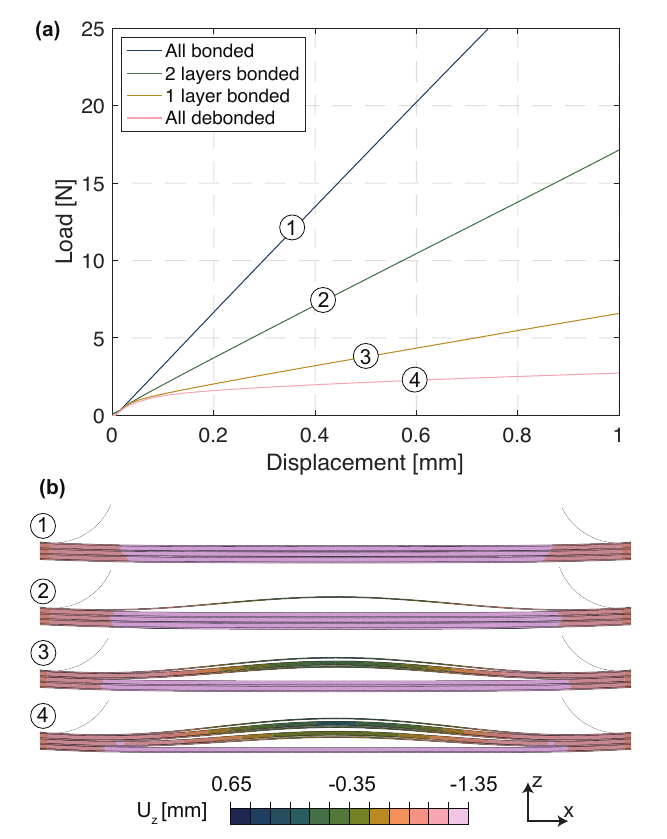}
	\caption{Control of stiffness ratio using 4 load-carrying layers. (a) Simulation results of force-displacement response of the hinge with $t_c/t_f = 3.31$. (b) Displacement ($u_z$) contours at $d = 1.0~mm$ with $t_c/t_f = 3.31$ for various bonded states. The numbers correspond to the curves in sub-figure $(a)$.}
	\label{fig:4Layer}
\end{figure}

We propose the expansion of the stiffness reprogrammability in the laminated hinges by repeating the dry adhesive and load-carrying layers. Figure~\ref{fig:3Layer}(a) and Figure~\ref{fig:4Layer}(a) show respectively the behavior of hinges with three and four load-carrying layers, each with $t_c/t_f = 3.31$. In both cases, we demonstrate that there are as many distinct stiffness layers as load-carrying layers in the hinges. Starting from all interfaces between the stiff layers and the dry adhesive layers being in the bonded state, we can debond the dry adhesive layers one at a time to achieve successively softer stiffness regimes. As long as the load-carrying layers are debonded sequentially starting from the compressive side of the bend, cooperative buckling between the debonded stiff layers is possible. This is shown in the simulated displacement contours in Figure~\ref{fig:3Layer}(c) and experimentally in Figure~\ref{fig:3Layer}(d) for a hinge with three load-carrying layers. Visually, there is good agreement between simulated and experimental deformation of the dry adhesive and load-carrying layers. Notably, all stiffness states exhibit a linear response with the exception of small displacements ($d < 0.1 mm$) where the buckling of the stiff layers occurs. The simulated bending response for the three layer hinge is experimentally verified and results are shown in Figure~\ref{fig:3Layer}(a), with stiffest-to-softest ratio of 14.4 and a intermediate-to-softest ratio of 5.2. For the four layer hinge, we note that a larger geometric imperfection of 60~$\mu m$ in simulation is required to achieve buckling in the softest state. The effect on bending stiffness of the increased imperfection is found to be negligible.

Figure~\ref{fig:3Layer}(b) shows the achievable stiffness ratios as a function of the thickness ratio for the three layer hinge. The stiffness ratios are always taken relative to the softest, fully debonded state. For example, the blue line shows the stiffness ratio of state 1 relative to 3 and the yellow line shows the stiffness ratio of state 2 relative to 3.

\section{Discussion}
\label{sec:discussion}
\subsection{Strategy for hinge design}
The optimal design strategy for the proposed reconfigurable hinges depends on the design objective. For a fixed total thickness, the stiffness variability is maximized by the design with two load-carrying layers. For example, consider a hinge with a total thickness of $t = 900 \mu m$ with $t_f = 100 \mu m$ load-carrying layers. The simulated stiffness ratio, $\rho$, for the two and three layer designs is 21.3 and 14.7 respectively. The elastic buckling of the debonded layers generally reduces the achieved stiffness modulation. On the other hand, the multi-layer hinges are advantageous for achieving more than two stiffness states or if there are limitations on the dry adhesive thickness that can be manufactured. Another consideration is that actuation becomes more difficult for thick dry adhesive layers since electrostatic force scales inversely with the square of the distance between the load-carrying layers \cite{Bergamini2007}.

\begin{table*}
    \footnotesize

	\begin{tabular}{P{0.20\linewidth}P{0.14\linewidth}P{0.14\linewidth}P{0.17\linewidth}P{0.14\linewidth}P{0.14\linewidth}}
		\textbf{Technology} & \textbf{Maximum number of stiffness states} & \textbf{Stiffness state maintainability} & \textbf{Shape change and stiffness modulation coupled?} & \textbf{Stiffness modulation ratio, $\rho$} & \textbf{References} \\ \hline
		
		Shape memory alloys & 2 & Active & No & 2 - 4 & \cite{Manti2016,MohdJani2014,Cross1969} \\
		
		Shape memory polymers & 3 & Active & No & 2 - 100 &  \cite{Behl2007,Firouzeh2017,MohdJani2014,McKnight2010,Chenal2014} \\
		
		Lamina jamming (electrostatic) & $\geq$ 4 & Active & No & 1.2 - 18 & \cite{Wang2019,DiLillo2013} \\
		
		Particle/lamina jamming (pneumatic) & 2 & Active & No & 3 - 50 & \cite{Steltz2009,Narang2018,Jiang2014,Wall2015} \\
		
		Tape springs & 2 & Active & Yes & \textless 400  & \cite{Pellegrino2015,Baek2020,Santer2017} \\
		
		Bi-stable tape springs & 2 & Passive & Yes & \textless 100 & \cite{Diaconu2007,Arrieta2014a,Jeon2011,Costantine2012} \\ \hline
		
		Gecko Inspired Laminates & $\geq$ 4 & Passive & No & 6.3 - 14.4 (up to 73 in simulation) & This Work \\ \hline
	\end{tabular}
    
	\caption{Comparison of achieved stiffness modulation in state of the art methods for mechanical adaptation in structural applications.}
    \label{tab:SOTARatios}
\end{table*}

\subsection{Comparison to existing methods}
Table \ref{tab:SOTARatios} summarizes key performance metrics for stiffness reprogramming methods in literature and this work. The main advantage of the proposed stiffness modulation method is that we can achieve multiple stiffness states and maintain each of them passively -- expanding the range of mechanical performance and improving energy efficiency. At the same time, the proposed method does not rely on a large geometric shape change to achieve stiffness modulation (unlike tape springs). In the Gecko-inspired hinge, all stiffness states are accessible at small displacements, while the hinge ends can be mounted with static boundary conditions. This allows it to be easily integrated into structures. 

In terms of stiffness modulation, our experimental results perform on par with laminar jamming methods, while easily surpassing the performance of shape memory alloys, used commonly for their variable stiffness performance as actuators. Simulations demonstrate that the stiffness modulation can be improved using simple geometric design parameters, placing the stiffness modulation on par with all other existing methods. 

\section{Conclusion}
\label{sec:conclusion} 
In this work, we proposed a Gecko-inspired hinge with dynamically reprogrammable stiffness achieved through reversible lamination of thin stiff layers using dry adhesives via electrostatic or magnetic stimuli. Reprogrammable hinges with up to four distinct stiffness states were studied. All stiffness states are maintained passively due to switchable bonding between load-carrying and dry adhesive layers. Actuation times on the order of $\sim$1s were found to be sufficient to activate bonding. The degree of stiffness modulation can be controlled through geometric parameters, with a quadratic proportionality to the ratio of adhesive to load-carrying layer thickness demonstrated. Experimental stiffness responses were found to be repeatable and could be accurately captured by FE simulations. Stiffness reprogramming by a factor of up to 14.4 was shown experimentally, with factors up to 73.0 promised by FE simulations. Hinges with two programmable stiffness states were demonstrated to have the largest stiffness modulation for a given total thickness. 

The novel concept addressed two key limitations of existing stiffness control approaches. Namely, all available stiffness states can be passively maintained with energy required only to switch between states and the stiffness reprogramming is independent of shape reconfiguration. The Gecko-inspired hinge reduces energy requirements in adaptive systems and has a broad range of applicability to stiffness modulation in both static (e.g., shape reconfiguration) and dynamic applications (e.g., control of vibration response). As such, the proposed concept is of utility in energy-constrained environments of space, remote robotic exploration, and biomedical devices.

A promising avenue for future work is modeling and predicting the adhesive strength. Improvement of the strength using micro-patterning of the dry adhesive or through modification of adhesive material would enable higher load-carrying capacity in the stiff state. Additionally, the development of extension dominated structures would broaden the applicability of the proposed concept. In this work, stiffness reprogramming is enabled in bending which is ideal for applications like deployable space structures, often packaged using bending of thin composites. Dry adhesion between load-carrying elements can also be applied to structures loaded in tension. In this case, turning adhesion on and off would allow surfaces to transmit loads or slide relative to each other, respectively. 

\section*{Data Availability}
All raw data required to reproduce the findings of this paper is available at: \href{https://doi.org/10.25740/zj277df0909}{https://doi.org/10.25740/zj277df0909}.

\section*{Acknowledgments}
This research was supported by startup funding from the School of Engineering at Stanford University. Kai Jun Chen was supported by a postgraduate fellowship from DSO National Laboratories - Singapore.

\section*{Author Contributions}
Kai Jun Chen - Experiment and simulations for electrostatic hinge, literature review, analysis, paper writing and editing. Maria Sakovsky - Idea conception, literature review, experiment and simulations for magnetic hinge, paper writing and editing.

\printbibliography

\end{document}